\documentclass[prd,a4paper,preprint,showpacs,byrevtex,onecolumn]
{revtex4}
\usepackage{graphicx}
\usepackage{dcolumn}
\usepackage{amsmath}
\usepackage{array}
\usepackage{bm}
\usepackage{textcomp}
\newcommand{\fr}{\frac}
\DeclareMathAlphabet{\mathbbm}{U}{bbm}{m}{n}
\SetMathAlphabet\mathbbm{bold}{U}{bbm}{bx}{n}
\begin{document}

\title{Baryonic mass formula in large $N_c$ QCD versus quark model}
\author{Claude \surname{Semay $^{a}$}}
\email[E-mail: ]{claude.semay@umh.ac.be}
\author{Fabien \surname{Buisseret $^{a}$}}
\email[E-mail: ]{fabien.buisseret@umh.ac.be}
\author{Nicolas \surname{Matagne $^{b}$}}
\email[E-mail: ]{nmatagne@ulg.ac.be}
\author{Florica \surname{Stancu} $^{b}$}
\email[E-mail: ]{fstancu@ulg.ac.be}
\affiliation{$^{a}$ Groupe de Physique Nucl\'{e}aire Th\'{e}orique,
Universit\'{e} de Mons-Hainaut,
Acad\'{e}mie universitaire Wallonie-Bruxelles,
Place du Parc 20, BE-7000 Mons, Belgium.\\
$^{b}$ University of Li\`ege, Institute of Physics B5, Sart Tilman,
B-4000 Li\`ege 1, Belgium.}

\date{\today}

\begin{abstract}
A connection is establisehd between the results of a potential quark
model with dynamical masses and the $1/N_c$ expansion mass formula used
in the description of baryon resonances. It is shown that a remarkable
compatibility exists between the two methods.
\end{abstract}




\maketitle

\section{Introduction}

So far the standard approach to baryon spectroscopy is the
constituent quark model where the Hamiltonian contains a spin
independent
part formed of the kinetic plus the confinement energies and a
spin dependent part given by a hyperfine interaction. The latter can be
either
due to one gluon exchange or to Goldstone boson exchange between
quarks, or it can be an
instanton induced interaction. The results are naturally model
dependent.

It is therefore very important to develop model independent methods that
can help in alternatively understanding baryon spectroscopy and that can
support
quark model assumptions. Large $N_c$ QCD offers such a method. In 1974
't Hooft
proposed to generalize QCD from SU(3) to SU($N_c$) \cite{HOOFT} where
$N_c$ is an arbitrary number of colors and suggested a perturbative
expansion
in the parameter $1/N_c$, applicable to all QCD regimes. Witten has
generalized
the approach to baryons \cite{WITTEN} and this
has lead to a powerful $1/N_c$ expansion method to study static
properties
of baryons, as for example, the masses, the magnetic moments, the
axial currents, etc. The method is systematic and predictive.
It is based on the discovery that, in the limit
$N_c \rightarrow \infty$, QCD possesses an exact contracted
SU(2$N_f$) symmetry  \cite{Gervais:1983wq,DM} where $N_f$ is the number
of flavors. This symmetry is
only approximate for finite $N_c$ so that corrections have to be added
in powers of $1/N_c$.
The $1/N_c$ expansion method has extensively and successfully been
applied
to ground state baryons \cite{DJM94,DJM95,Jenkins:1998wy,EJ2} (for
recent developments
see Ref.~\cite{TRENTO}).
Its applicability to excited states is a subject of current
investigations.
In this case the symmetry under consideration
is assumed to be SU(2$N_f$) $\times$ O(3) where
SU(2$N_f$) is related QCD, as introduced above. However
O(3) is not related to QCD but it
brings an additional degree of freedom. It is of common practice to
introduce it in order to construct orbitally excited states. The direct
product SU(2$N_f$) $\times$ O(3) is also used in quark models to
classify
three quark states, but there SU(2$N_f$) is not an intrinsic symmetry.
Thus the two approaches have formally the same symmetry in common
which does not imply common dynamical assumptions. The only common
feature is that the excited states are stable in a first approximation.

The purpose of the present study is to see whether or not there is
a compatibility between the two methods. If such a compatibility exists,
an important support to the constituent quark model can be provided
by the model independent $1/N_c$ expansion method, and a better
understanding of the physical content of large $N_c$ mass formulas can
be gained.

In the language of quark models, the baryon states can roughly be
classified into excitation bands with $N = 0$ for the ground state band
and $N = 1,2,3,\ldots$ for excited states, where $N$ represents units
of excitation, like in a harmonic oscillator picture.
The key tool of this comparative study is that
one can analyze both the
$1/N_c$ expansion results and the quark model basic ingredients in terms
of $N$ which makes the comparison between the two methods possible and
very convenient.

The paper is organized as follows. The next section introduces the mass
formula used in the $1/N_c$ expansion method.
Section III gives a mass formula obtained from a Hamiltonian quark
model where the kinetic energy is relativistic, the confinement
is an Y-junction flux tubes and the
hyperfine interaction is of an one-gluon exchange nature.
Section IV is devoted to the comparison between terms of the mass
formula which are common in the two approaches. The last section is
devoted to conclusions.

\section{Baryons in large $N_c$ QCD}

For simplicity, we illustrate the method with the $N_f$ = 2 case
but the arguments are similar to any $N_f$.
So, here we deal with SU(4) which has 15 generators, the spin subgroup
generators
$S_i$ ($i = 1,2,3$), the isospin subgroup generators
$T_a$ ($a = 1,2,3$) and $G_{ia}$ which act both on spin and isospin
degrees of freedom. The SU(4) generators
are components of an irreducible tensor operator
which transforms according to the adjoint representation $[211]$ of
dimension $\bf 15$ of SU(4).
The SU(4) algebra is
\begin{eqnarray}\label{ALGEBRASU4}
&[S_i,T_a] = 0,
~~~~~ [S_i,G_{ja}]  =  i \varepsilon_{ijk} G_{ka},
~~~~~ [T_a,G_{ib}]  =  i \varepsilon_{abc} G_{ic},\nonumber \\
&[S_i,S_j]  =  i \varepsilon_{ijk} S_k,
~~~~~ [T_a,T_b]  =  i \varepsilon_{abc} T_c,\nonumber \\
&[G_{ia},G_{jb}] = \fr{i}{4} \delta_{ij} \varepsilon_{abc} T_c
+\fr{i}{4} \delta_{ab}\varepsilon_{ijk}S_k.
\end{eqnarray}
Together with the generators $\ell_i$ of SO(3), the SU(4)
generators form the building blocks of the mass operator.
Then in the $1/N_c$ expansion the mass operator $M$ has the general
form
\begin{equation}
\label{massoperator}
M = \sum_{i} c_i O_i,
\end{equation}
where the coefficients $c_i$ are reduced matrix elements that
encode the QCD dynamics and are
determined from a fit to the existing data, and the operators $O_i$
are O(3) scalars of the form
\begin{equation}\label{OLFS}
O_i = \frac{1}{N^{n-1}_c} O^{(k)}_{\ell} \cdot O^{(k)}_{SF},
\end{equation}
where  $O^{(k)}_{\ell}$ is a $k$-rank tensor in O(3) and $O^{(k)}_{SF}$
a $k$-rank tensor in SU(2)-spin (homomorphic to SO(3)),
but invariant in SU(2)-flavor. Generally
the operators $O^{(k)}_{SF}$ are combinations
of the SU(2$N_f$) generators and here, in particular, of SU(4)
generators.
The lower index $i$ in the left hand side represents a specific
combination.
Each $n$-body operator is multiplied by an explicit factor of
$1/N^{n-1}_c$ resulting from the power counting rules \cite{WITTEN}.
For the ground state, one has $k$ = 0. For excited states the $k = 2$
tensor is also important. The sum in the mass operator is finite.
Operator reduction rules simplify the expansion.
In addition, in practical applications, it is customary to include terms
up to $1/N_c$ and drop higher order corrections of order $1/N_c^2$.
As an example, in Eqs.~(\ref{example}), 
we exhibit the list of operators
used in the calculation of the masses of the $[{\bf 70},1^-]$ multiplet
up to
order $1/N_c$ included \cite{Matagne:2006dj}. Note that although
$O_5$ and $O_6$ carry a factor of $1/N_c^2$ their matrix elements are
of order $1/N_c$ because they contain the coherent operator $G^{ia}$
which brings an extra factor of $N_c$.
\begin{eqnarray}\label{example}
O_1 = N_c \, \openone \negthinspace \negthinspace\negthinspace
\negthinspace\negthinspace&, &
~O_2  =  \frac{1}{N_c}\ell^i S^i,
~~~O_3 = \frac{1}{N_c} T^aT^a,
~~~O_4 = \frac{1}{N_c} S^iS^i,\nonumber\\
& O_5 & =  \frac{15}{N_c^2}\ell^{(2)ij}G^{ia}G^{ja},
~~~O_6 =  \frac{3}{N_c^2}\ell^iT^aG^{ia}.
\end{eqnarray}
Here $O_1 = N_c \, \openone $ is the trivial operator, proportional
to
$N_c$ and the only one which survives when $N_c \rightarrow \infty$
\cite{WITTEN}, where the SU(4) symmetry is exact. It is the only
spin-isospin
independent term in the mass formula. The SU(4) quadratic operators
$S^iS^i$, $T^aT^a$ and $G^{ia} G^{ia}$ should all enter the mass
formula (the sum over repeated indices is implicit). But they are
related to each other by the operator identity~\cite{Jenkins:1998wy}
\begin{equation}
\label{CASIMIR}
 \left\{S^i,S^i\right\} + \left\{T^a,T^a\right\} + 4  \left\{G^{ia},
 G^{ia}\right\} = \frac{1}{2} N_c (3N_c + 4),
\end{equation}
so one can express $G^{ia} G^{ia}$ in terms of $S^iS^i$ and $T^aT^a$.
Note that the right hand side of Eq.~(\ref{CASIMIR}) is the eigenvalue
of the Casimir operator for the irreducible representation $[N_c - 1,1]$
of SU(4).
The operators $O_2$, $O_5$ and $O_6$ are relevant for orbitally excited
states.
Among them, the role of $O_2$ will be discussed below.

\subsection{The ground state band}

The mass formula for the ground state up to order $1/N_c$ is
simple because one can replace $T^aT^a$ by $S^iS^i$, due to an
identity which holds for symmetric $[N_c]$ states
\cite{Jenkins:1998wy}.
As there is no orbital excitation, the mass formula~(\ref{massoperator})
takes the following simple form
\begin{equation}\label{GS}
 M = c_1 N_c + c_4\frac{1}{N_c}S^2 + \mathcal{O}\left(\frac{1}{N_c^3}
 \right),
\end{equation}
which means that
for $N = 0$ only the operators $O_1$ and $O_4$ contribute to the
mass. Thus the fit gives quantitative information only for $c_1$ and
$c_4$.
For $N_c = 3$, $M_N = 940$~MeV for $S=1/2$, and
$M_{\Delta} = 1232$~MeV for $S=3/2$, one gets
\begin{equation}\label{C2C4}
c_1 = 289 ~ \mathrm{MeV},~~~~~   c_4 = 292 ~ \mathrm{MeV}.
\end{equation}

\subsection{Excited states}

Among the excited states, those belonging to the $N = 1$ band,
or equivalently to the $[{\bf 70},1^-]$ multiplet,
have been most extensively studied,
either for $N_f = 2$
\cite{CGKM,Goi97,PY1,PY2,CCGL,CaCa98,BCCG,Pirjol:2003ye,COLEB,cohen1}
or for $N_f = 3$ \cite{SGS}. In the latter case,
first order corrections in SU(3) symmetry breaking
were also included.

The $N = 2$ band contains the $[{\bf 56'},0^+]$, $[{\bf 56},2^+]$,
$[{\bf 70},\ell^+]$ ($\ell$ = 0, 2) and $[{\bf 20}, 1^+]$ multiplets.
There are no physical resonances associated to $[{\bf 20}, 1^+]$.
The few studies related to the $N = 2$ band concern the
$[{\bf 56'},0^+]$ for $N_f$ = 2 \cite{CC00}, $[{\bf 56},2^+]$
for $N_f = 3$ \cite{GSS}, and $[{\bf 70},\ell^+]$ for
$N_f = 2$ \cite{MS2}, later extended to $N_f = 3$
\cite{Matagne:2006zf}.
The method has also been applied \cite{MS1}
to highly excited non-strange and strange
baryons belonging to $[{\bf 56},4^+]$,
the lowest multiplet of the $N = 4$ band \cite{SS94}.

The group theoretical similarity of excited symmetric states
to the ground state makes the analysis of these states simple
\cite{GSS,MS1}.
For mixed symmetric states, the situation is more complex.
There is a standard procedure which reduces the study of mixed
symmetric states to that of symmetric states.
This is achieved by the decoupling of the baryon into an excited
quark and a symmetric core of $N_c - 1$ quarks.
This procedure has been applied to the $[{\bf 70},1^-]$ multiplet
\cite{
CGKM,Goi97,PY1,PY2,CCGL,CaCa98,BCCG,Pirjol:2003ye,COLEB,cohen1,SGS}
and to the $[{\bf 70},\ell^+]$ ($\ell$ = 0, 2) multiplet
\cite{MS2,Matagne:2006zf}. In fact
the decoupling is not necessary, provided
one knows the matrix elements of the SU(2$N_f$) generators
between mixed symmetric states. The case of SU(4) has been presented
in Ref. \ \cite{Matagne:2006dj}.

In Section IV, we collect the values of $c_1$, $c_2$ and $c_4$ obtained
in
the above studies in order to make a comparison between those values
and their analogs resulting from the quark model described below.

\section{Quark model for baryons}

\subsection{Confining interaction}

In the framework of potential models, it is generally assumed that a
baryon, viewed as a bound state of three
quarks, can be described in a first approximation by the following
spinless Salpeter Hamiltonian
\begin{equation}\label{ssh}
  H=\sum^3_{i=1}\sqrt{\vec p^{\, 2}_i+m^2_i}+V_Y,
\end{equation}
where $m_i$ is the current mass of the quark $i$, and $V_Y$ the
confining interaction potential.

The nonperturbative part of the gluon exchanges, responsible for the
confinement, can be successfully described in the flux tube model
\cite{CKP}. In
this framework, each quark is assumed to generate a string, or a flux
tube, characterized by its energy density (string tension).
Recent developments in lattice QCD tend to confirm the Y-junction as the
correct configuration for the flux tubes in
baryons \cite{Koma}. In this picture, a flux tube starts from each quark
and the tubes meet at the Toricelli point of the triangle formed by
the three quarks. This point, denoted by $\vec x_{T}$, is such that it
minimizes
the sum of the flux tube lengths, and its position is a complicated
function of the quark coordinates $\vec{x}_{i}$. Moreover, the energy
density of the tubes appears to be equal for mesons
and baryons. The Y-junction potential reads
\begin{equation}
V_Y=a \sum^{3}_{i=1} \left|\vec{x}_{i}-\vec{x}_{T}\right|.
\end{equation}
In Ref.~\cite{Bsb04}, it has been shown that this complicated potential
is successfully approximated by the more easily computable expression
\begin{equation}\label{pot1}
V=a\left[\alpha \sum^{3}_{i=1}\left|\vec{x}_{i}-\vec{R}\right| +(1-
\alpha)\frac{1}{2}\sum_{i<j}\left|\vec{x}_{i}-\vec{x}_{j}\right|\right],
\end{equation}
where $\vec{R}$ is the position of the center of mass. If $\alpha=1$,
Eq.~(\ref{pot1}) is a simplified Y-junction, where the Toricelli point
is replaced by the center of mass. If $\alpha=0$, this interaction
reduces
to a $\Delta$-type potential. Results of
Ref.~\cite{Bsb04}, obtained in the framework of a potential model, show
that $\alpha=1$ gives a better description than $\alpha=0$, and that the
Y-junction is approximated at best by $\alpha$ close to $1/2$.

\subsection{Mass formula}

Let us now introduce auxiliary fields, in order to get rid of the square
roots appearing in the Hamiltonian (\ref{ssh}). We get
\begin{eqnarray}\label{ham3b}
H(\mu_i,\nu_j,\lambda_{ij})&=&\sum^3_{j=1}\left[\frac{\vec{p}^{\, 2}_j+m
^2_j}{2\mu_j}+\frac{\mu_j}{2}\right]\nonumber\\
&&+\alpha\sum^3_{j=1}\left[\frac{ a^2 (\vec{x}_j-\vec{R})^2}{2\nu_j}+
\frac{\nu_j}{2}\right]+\frac{(1-\alpha)}{2}\sum_{j<k}\left[\frac{a^2 (
\vec{x}_j-\vec{x}_k)^2}{2\lambda_{jk}}+\frac{\lambda_{jk}}{2}\right].
\end{eqnarray}
The auxiliary fields, denoted as $\mu_i,\, \nu_j$, and $\lambda_{ij}$
are, strictly speaking, operators. Although being formally simpler,
$H(\mu_i,\nu_j,\lambda_{ij})$ is equivalent to $H$ up to the elimination
of the auxiliary fields thanks to the constraints
\begin{subequations}\label{elim}
\begin{eqnarray}
  \delta_{\mu_i}H(\mu_i,\nu_j,\lambda_{ij})&=&0\ \Rightarrow\ \mu_{i,0}=
  \sqrt{\vec{p}^{\, 2}_i+m^2_i},\\
  \delta_{\nu_j}H(\mu_i,\nu_j,\lambda_{ij})&=&0\ \Rightarrow\ \nu_{i,0}=
  a|\vec{x}_i-\vec{R}|,\\
  \delta_{\lambda_{ij}}H(\mu_i,\nu_j,\lambda_{ij})&=&0\ \Rightarrow\
  \lambda_{ij,0}=a|\vec{x}_i-\vec{x}_j|.
\end{eqnarray}
\end{subequations}
It is worth mentioning that $\left\langle \mu_{i,0}\right\rangle$ can be
seen as a dynamical mass of a quark of current mass $m_i$, while
$\left\langle \nu_{i,0}\right\rangle$ is, in this case, the static
energy
of the straight string linking the quark $i$ to the Toricelli
point~\cite{Fab1}. Similarly, $\left\langle \lambda_{ij,0}\right\rangle$
can be interpreted as the static energy of a straight string joining the
quarks $i$ and $j$. Although the auxiliary fields are operators, the
calculations are considerably
simplified if one considers them as real numbers. They are then finally
eliminated by a minimization of the masses with respect to
them \cite{Sem03}. The extremal values of $\mu_i$, $\nu_j$, and
$\lambda_{ij}$, considered
as numbers,
are logically close to the values of
$\left\langle \mu_{i,0}\right\rangle$,
$\left\langle \nu_{j,0}\right\rangle$, and
$\left\langle\lambda_{ij,0}\right\rangle$ given by
relations~(\ref{elim}). This procedure leads to a spectrum which is an
upper bound of the ``true spectrum" (computed without auxiliary fields)
\cite{Fab3}:
it can be shown that, the more auxiliary fields are introduced, the
higher
are the masses compared to those without auxiliary fields \cite{hyb1}.
Let us finally mention that, for $\alpha=1$, the
Hamiltonian~(\ref{ham3b})
can be related to the rotating string model for a baryon (see for
example Ref.~\cite{Simo2}).

In Ref.~\cite{coqm}, it has been shown that the eigenvalues of a
Hamiltonian of the form (\ref{ham3b}) can be analytically found by
making an appropriate change of variables, the quark coordinates
$\vec x_{i}=\left\{\vec x_{1},\vec x_{2},\vec x_{3}\right\}$ being
replaced by new coordinates
$\vec x'_{k}=\left\{\vec R,\vec \xi,\vec \eta\right\}$. The center
of mass is defined as
\begin{equation}\label{cmdef}
\vec R=\frac{\mu_{1}\vec x_{1}+\mu_{2}\vec x_{2}+\mu_{3}\vec x_{3}}{\mu_
{t}},
\end{equation}
with $\mu_{t}=\mu_{1}+\mu_{2}+\mu_{3}$ and $\{\vec \xi,\vec \eta\}$
being the
two relative
coordinates. From Ref.~\cite{coqm}, it can be immediately found that the
mass
spectrum of bound states of three massless particles ($m_i=0$ for the
$u$ and $d$ quarks) is given by
\begin{equation}\label{mass3m}
M(\mu,\nu,\lambda)=\omega (2n+\ell+3)+\frac{3}{2}\left(\mu+\alpha\nu+
\frac{(1-\alpha)}{2}\lambda\right),
\end{equation}
with
\begin{equation}
  \omega=a\sqrt{\frac{1}{\mu}\left[\frac{\alpha}{\nu}+\frac{3(1-\alpha)}
  {2\lambda}\right]},
\end{equation}
$n=n_\xi+n_\eta$ and $\ell=\ell_\xi+\ell_\eta$.
An obvious symmetry argument helps us to make the identification
$\mu_i=\mu$,
$\nu_i=\nu$, and $\lambda_{ij}=\lambda$.
In this symmetric case, properties of the equilateral triangle
together with the relations~(\ref{elim}) allow to make the following
ansatz
\begin{equation}
  \lambda=\sqrt{3}\, \nu.
\end{equation}
Defining
\begin{equation}
  Q=\alpha+\frac{(1-\alpha)}{2}\sqrt{3},\quad \tilde{\nu}=Q\nu,
\end{equation}
and
\begin{equation}\label{N}
  N= 2 n+\ell,
\end{equation}
we find
\begin{equation}\label{mass3_s}
  M(\mu,\tilde{\nu})=aQ\sqrt{\frac{1}{\mu\tilde{\nu}}}(N+3)+\frac{3}{2}
  \left(\mu+\tilde{\nu}\right).
\end{equation}
Formula (\ref{mass3_s}) is clearly symmetric in $\mu$ and
$\tilde{\nu}$. That means that we can set $\mu=\tilde{\nu}$.
This equality can be viewed as a sort of virial theorem. Then we have
\begin{equation}
  M(\mu)=\frac{aQ}{\mu}(N+3)+3\mu.
\end{equation}
One can easily find that the relation $\delta_\mu M(\mu)=0$ implies
\begin{equation}
  \mu_0=\left[\frac{a}{3}Q(N+3)\right]^{1/2},
\end{equation}
and $M(\mu_0)=6\mu_0$, as observed in Ref.~\cite{Simo3}. Writing
explicitly the square mass, we see that the model of Ref.~\cite{Simo3}
also predicts Regge
trajectories, which are in agreement with the experimental data
for light baryons
\begin{equation}\label{mass3q}
M^2(\mu_0)=12\, a\, Q\, (N+3).
\end{equation}
The Regge slope is here given by $12aQ$. However, from experiment we
know that the Regge slope for light baryons and light mesons are
approximately equal. For light mesons, the exact value in the
relativistic flux tube model is $2\pi a$, a
lower value than the one obtained from formula~(\ref{mass3q}). This is
due
to the auxiliary fields method: the more auxiliary fields we introduce,
the more the masses are overestimated \cite{hyb1}. What can be done to
cure this problem is to rescale $a$: let us define $\sigma$ such that
$12aQ=2\pi \sigma$. Then, formula~(\ref{mass3q}) is able to reproduce
the light baryon Regge slope for a physical value $\sigma\lesssim 0.2$
GeV$^2$. Note that the best value for $\alpha$ is $1/2$. Consequently,
the best value for $Q$ is $1/2+\sqrt{3}/4\approx0.93$. It is worth
mentioning that such a rescaling of the string tension has already given
good results in the study of hybrid mesons \cite{hyb2}.

\subsection{One gluon exchange and quark self-energy}

Although including only the confining energy is sufficient to
understand the Regge trajectories of light baryons, it is well-known
that the absolute value of the masses which are obtained are too high
with respect to the experimental data. Other contributions are needed to
decrease these masses and we shall estimate their effect
perturbatively. The most widely
used is a Coulomb interaction term of the form
\begin{equation}
  \Delta M_{oge}=-\frac{2}{3}\alpha_s\sum_{i<j}\left\langle \frac{1}{|
  \vec x_i-\vec x_j|}\right\rangle,
\end{equation}
arising from one gluon exchange processes, where $\alpha_s$ is the
strong
coupling constant, usually assumed to be around $0.4$ for light hadrons
\cite{scc}. Lattice QCD calculations also support this value
\cite{lat0}. Assuming that
$\left\langle 1/A\right\rangle\approx1/\left\langle A\right\rangle$, and
using symmetry arguments, relations~(\ref{elim}) lead to
\begin{equation}
  \sum_{i<j}\left\langle \frac{1}{|\vec x_i-\vec x_j|}\right\rangle
  \approx\frac{3\, a}{\lambda_0}=\frac{\sqrt{3}\, a\, Q}{\mu_0},
\end{equation}
\begin{equation}
  \Delta M_{oge}=-2\alpha_s\frac{a\, Q}{\sqrt{3}\, \mu_0}.
\end{equation}

Another interesting contribution to the mass, which can be added
perturbatively, is the quark-self energy. Recently, it was shown
that the quark self-energy, which is created by the color
magnetic moment of a quark propagating through the vacuum background
field, adds a negative constant to the
hadron masses \cite{qse}. Its negative sign is due to the paramagnetic
nature of the particular mechanism at work in
this case. The quark self-energy contribution for three massless quarks
is given by \cite{qse}
\begin{equation}
  \Delta M_{qse}=-\frac{3fa}{2\pi\mu_0}.
\end{equation}
The factor $f$ has been computed in lattice QCD studies. First quenched
calculations gave $f = 4$ \cite{qse2}.
A more recent unquenched work \cite{qse3} gives $f = 3$. Since its value
is still a matter of research, we will only assume that $f\in[3,4]$.

With the unperturbed baryon
mass $M(\mu_0)$, given by Eq.~(\ref{mass3q}), the total mass is given by
the sum
$M_0 = M(\mu_0)+\Delta M_{oge}+\Delta M_{qse}$.
Then, in the first order of perturbation and for
$\alpha=1/2$, it is straightforward to obtain
the following mass formula for baryons
\begin{equation}
\label{massf}
  M^2_0 =
  2\pi\sigma(N+3)-\frac{4}{\sqrt{3}}\pi\sigma\alpha_s-\frac{12}{(2+
  \sqrt{3})}f\sigma,
\end{equation}
where the scaling $12aQ=2\pi \sigma$ has been used.
The effects of the one gluon exchange term and of the quark self-energy
are thus to shift the square mass spectrum by a global negative
amount. Let us note that the symbol $N$
defined by Eq.~(\ref{N}) and the
quantity $N$ used to classify baryon states and used to plot results
from
the $1/N_c$ expansion are the same. This common $N$ will be used in the
next section to perform a comparison between the results obtained in
both approaches.

The mass formula~(\ref{massf}) does not take into account spin
relativistic
contributions, as the spin-spin or spin-orbit forces.
Within the auxiliary field formalism, all these
corrections to the static potential are expanded in powers
of $1/\mu^2$ where $\mu$ is the constituent quark mass \cite{morg99}.
All the spin corrections to the mass formula~(\ref{massf}) must depend
both
on the matrix elements of the interaction and on the coefficient
$1/\mu^2$. In the following, we shall consider that the dominant
dynamical effect is due to the constituent mass, while the matrix
elements remain roughly constant with $N$, as presented in the next
section.

\section{Comparison of the two approaches}

\begin{figure}[ht]
\includegraphics[width=9cm]{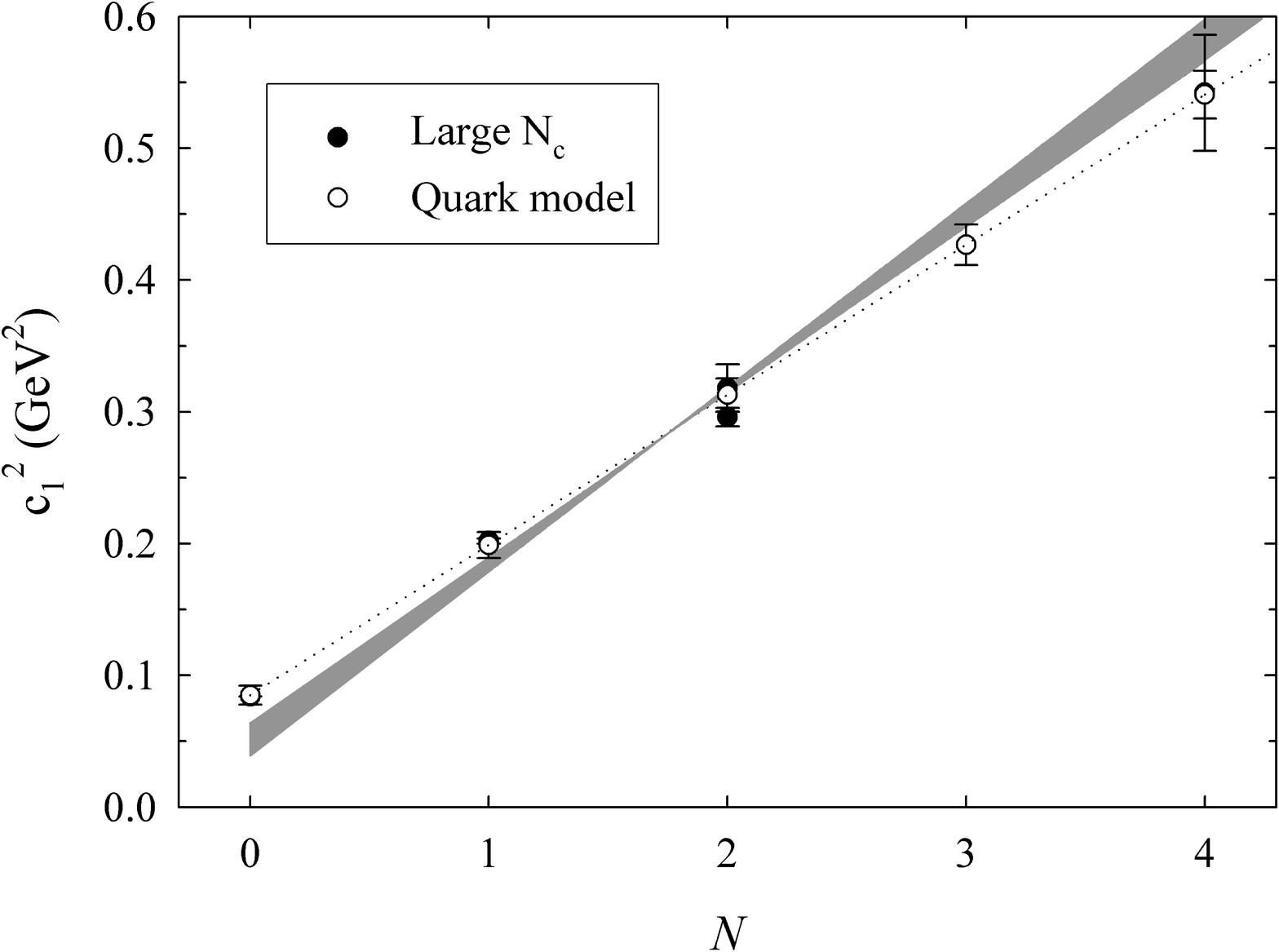}
\caption{Values of $c^2_1$ computed in the $1/N_c$ expansion (full
circles) from a fit to
experimental data
(Eq.~(\ref{C2C4}) for $N = 0$, Refs.~\cite{SGS,GSS} for $N = 1$,
Ref. \cite{MS2} for $N = 2$ and Ref. \cite{MS1} for
$N = 4$), compared with results from a fit (see text) of the
formula~(\ref{c1qm})
(empty circles and dotted line to guide the eyes). No data is available
for $N=3$ in large $N_c$ studies. Values of $c^2_1$ as predicted by
formula (\ref{c1qm}) for $\sigma\in[0.17,0.20]$~GeV$^2$ are located in
the shaded area.}
\label{Fig1}
\end{figure}

In the $1/N_c$ expansion method, the first term $c_1 N_c$ in the mass
formula
of Eq.~(\ref{massoperator}) contains the main
spin-independent contribution to the baryon mass, which in a
quark model language, represents the confinement and the
kinetic energy. So, it is natural to identify this term with the
mass given by the formula~(\ref{massf}). Then, for $N_c = 3$, we assume
the relation
\begin{equation}
c^2_1=\frac{M^2_0}{9},
\end{equation}
 which gives
\begin{eqnarray}\label{c1qm}
  c^2_1&=&\frac{2\pi}{9}\sigma N+c_0\\
  &=&\frac{2\pi}{9}\sigma(N+3)-\frac{4}{9\sqrt{3}}\pi\sigma\alpha_s-
  \frac{4}{3(2+\sqrt{3})}f\sigma.
\end{eqnarray}
Fig.~\ref{Fig1} shows a comparison between the values of $c^2_1$
obtained in the $1/N_c$ expansion method and those derived from the
Eq.~(\ref{c1qm}) for various values of $N$. From this comparison
one can see that the results of large $N_c$ QCD are entirely
compatible with the formula~(\ref{c1qm}). From a fit, one has
$\sigma=0.163\pm0.004$~GeV$^2$, a rather low but still acceptable value
according to usual potential models,
and $c_0=0.085\pm0.007$~GeV$^2$. To reproduce $c_0$, we can set
$\alpha_s=0.4$, $f=3.5$: these are very standard values.

In most of the quark models however, the string tension is generally
assumed to lie in the range $[0.17,0.20]$~GeV$^2$. If the value of
$\sigma$ is chosen in this interval, the corresponding values for
$c^2_1$, given by Eq.~(\ref{c1qm}), are located in the shaded area of
Fig.~\ref{Fig1}. Although the agreement with large $N_c$ data is not so
good than in the optimal case, where $\sigma=0.163$~GeV$^2$, it remains
satisfactory if we choose $f=3.98\ (4.42)$ for
$\sigma=0.17\ (0.20)$~GeV$^2$, together with $\alpha_s=0.4$.
These values are larger than what
is expected. It could be argued
that other mechanisms than the quark self-energy are present, their
contribution decreasing the total mass $M_0$. In mesons for example,
retardation effects due to the finite interaction speed were shown to be
also proportional to $\mu^{-2}$, like the quark self-energy
\cite{Buisret}. It is possible that, when retardation effects are
included,
$f$ can again be chosen in the interval $[3,4]$ with a standard value of
$\sigma$. But, no model for retardation effects in baryon has been
proposed yet.
\begin{figure}[ht]
\includegraphics[width=9cm]{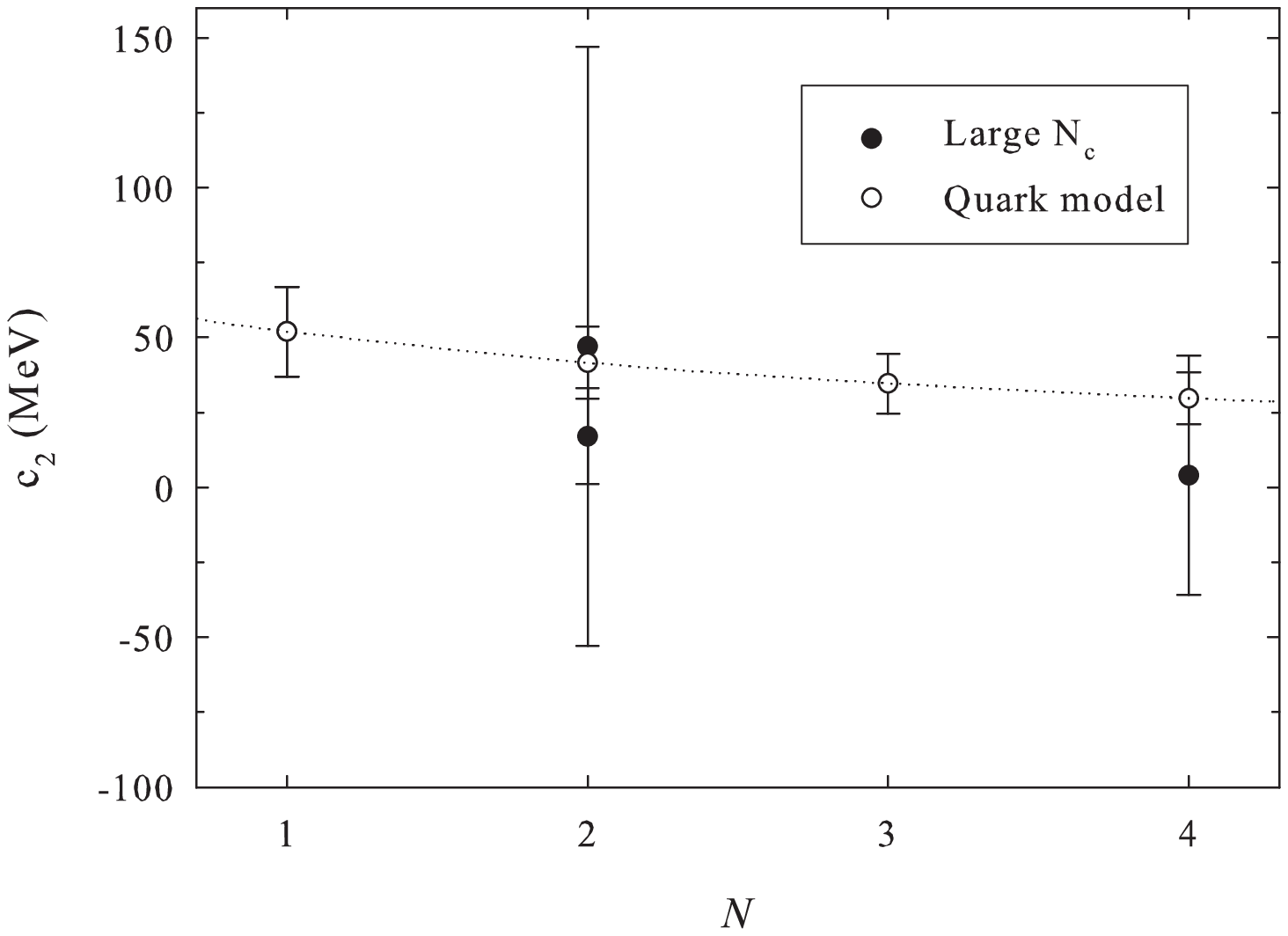}
\caption{
Values of $c_2$ computed in the $1/N_c$ expansion (full circles) from a
fit to
experimental data (Refs.~\cite{SGS,GSS} for $N = 1$, Ref. \cite{MS2}
for $N = 2$ and Ref. \cite{MS1} for $N = 4$), compared with results
from formula~(\ref{c2c4qm})
(empty circles and dotted line to guide the eyes). No data is available
for $N=3$ in large $N_c$ studies.}
\label{Fig2}
\end{figure}

\begin{figure}[ht]
\includegraphics[width=9cm]{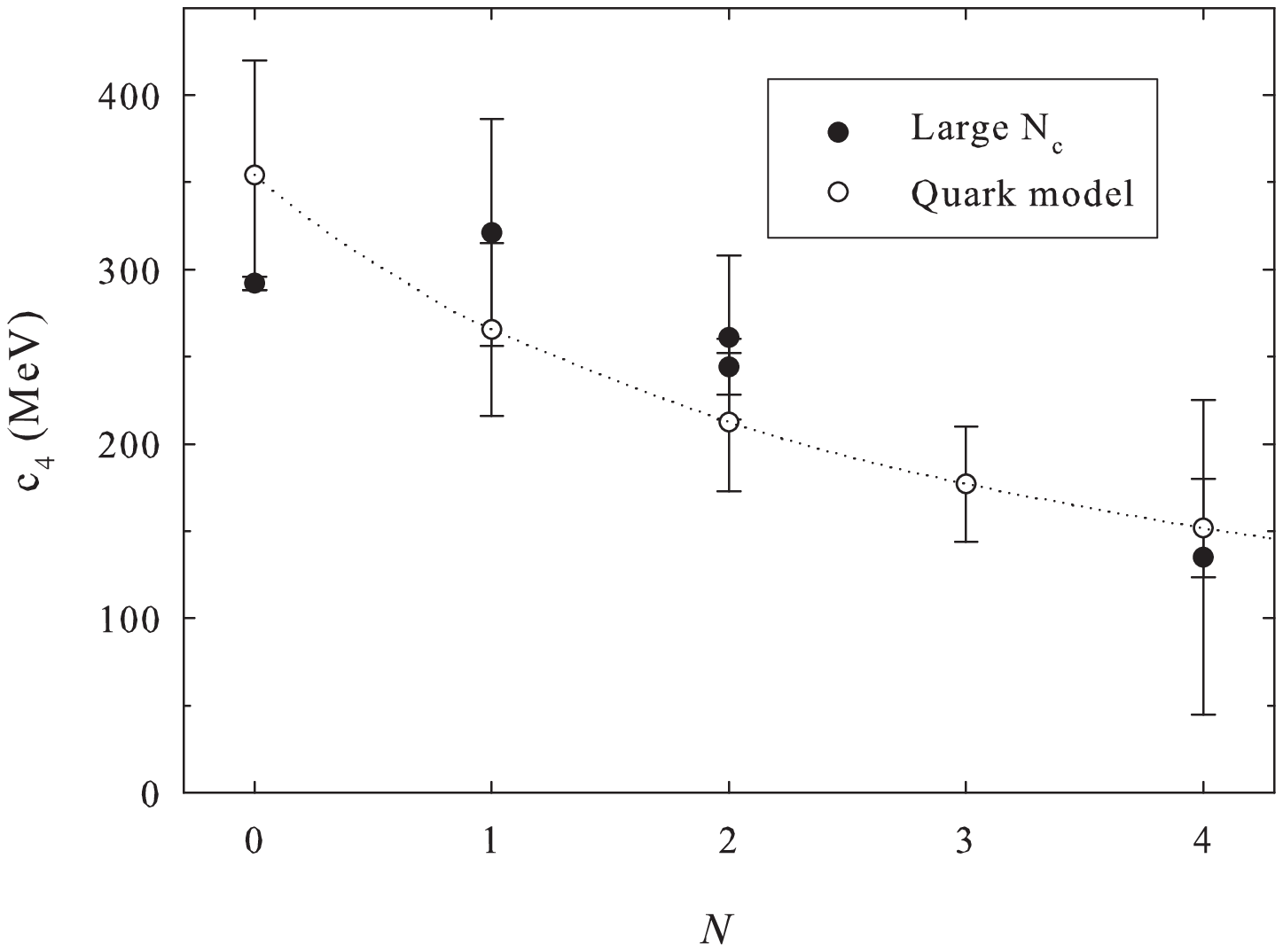}
\caption{
Values of $c_4$ computed in the $1/N_c$ expansion (full circles) from a
fit to
experimental data
(Eq.~(\ref{C2C4}) for $N = 0$, Refs.~\cite{SGS,GSS} for $N = 1$,
Ref. \cite{MS2} for $N = 2$ and Ref. \cite{MS1} for
$N = 4$), compared with results from formula~(\ref{c2c4qm})
(empty circles and dotted line to guide the eyes). No data is available
for $N=3$ in large $N_c$ studies.}
\label{Fig3}
\end{figure}

Within the auxiliary field formalism, we can expect that
$c_2$ and $c_4\propto\mu^{-2}_0$, and thus
\begin{equation}\label{c2c4qm}
c_2=\frac{c^0_2}{N+3},\quad c_4=\frac{c^0_4}{N+3}.
\end{equation}
We see that this behavior is coherent with the large $N_c$ results in
Figs.~\ref{Fig2} and \ref{Fig3}. We chose $c^0_2=208\pm60$~MeV so that
the point with $N=1$, for which the uncertainty is minimal, is exactly
reproduced. Let us note that the spin-orbit term is vanishing for $N=0$,
so no large $N_c$ result is available in this case. To compute the
parameter $c^0_4$ a fit is performed on all the
large $N_c$ data. We obtain then $c^0_4=1062\pm198$~MeV.
Note that $c^0_4\gg c^0_2$. This shows that the spin-spin contribution
is much larger than the spin-orbit contribution, which justifies
the neglect of the spin-orbit one in quark model
studies.

\section{Conclusions}

This study supports the quark model basic assumptions by
the compatibility of its mass formula with the mass formula
derived from the model independent
$1/N_c$ expansion. These assumptions are:
 relativistic kinetic energy for light quarks,
Y-junction confining interaction, negligible spin-orbit
interaction, hyperfine interaction dominated by a spin-spin term.
A recent analysis shows that a flux tube model
and a feeble spin-orbit interaction give a succesful account
of hadron spectroscopy \ \cite{WILCZEK}.

In addition this study
suggests that a good description of the bulk content
of the baryon mass can be obtained with a spin independent energy
eigenvalue
of the form $M_0 \propto \sqrt{N + 3}$ where $N = 0,1,2,\ldots$ is the
number of excitation units, as in the harmonic oscillator.
It also shows that the spin-orbit and spin-spin interactions vanish with
the excitation energy. Moreover this comparative
study gives a better insight into the large $N_c$ mass operator
where the coefficients $c_i$ encode the QCD dynamics.

\acknowledgments
Financial support is acknowledged by C.S. and F.B. from FNRS (Belgium)
and by N.M. from IISN (Belgium).


\end{document}